\documentclass[aps,twocolumn]{revtex4}
\usepackage{graphicx,amsmath,amsfonts,amssymb}

\begin{document}

\title{Learning-rate dependent clustering and self-development in a network of coupled phase oscillators}
\author{Ritwik K. Niyogi$^{1,2}$ and L. Q. English$^{1}$ }
\affiliation{$^{1}$Department of Physics and Astronomy \\ $^{2}$Program in Neuroscience and Department of Mathematics \\ Dickinson
College, Carlisle, Pennsylvania, 17013, USA}
\date{\today}

\begin{abstract}
We investigate the role of the learning rate in a Kuramoto Model of coupled phase oscillators in which the coupling coefficients dynamically vary according to a Hebbian learning rule. According to the Hebbian theory, a synapse between two neurons is strengthened if they are simultaneously co-active. Two stable synchronized clusters in anti-phase emerge when the learning rate is larger than a critical value. In such a fast learning scenario, the network eventually constructs itself into an all-to-all coupled structure, regardless of initial conditions in connectivity. In contrast, when learning is slower than this critical value, only a single synchronized cluster can develop. Extending our analysis, we explore whether self-development of neuronal networks can be achieved through an interaction between spontaneous neural synchronization and Hebbian learning. We find that self-development of such neural systems is impossible if learning is too slow. Finally, we demonstrate that similar to the acquisition and consolidation of long-term memory, this network is capable of generating and remembering stable patterns.
\end{abstract}

\maketitle

\section{Introduction}
\label{sec:intro}

Spontaneous mass-synchronization has been observed in several biological systems, such as in the synchronous flashing of fireflies \cite{Buck, BuckBuck}, the chirping of crickets \cite{walker}, and in the pacemaker cells in the cardiovascular system \cite{michaels}. Within the nervous system, synchronous clustering has been reported in networks of neurons in the visual cortex \cite{Sompolinsky1}, olfactory bulb \cite{Galanetal}, central pattern generators \cite{KopellErmentrout,SigvardtWilliams,RandCohenHolmes,CruzCortez} as well as in those involved in generating circadian rhythms \cite{liu}. Neuronal synchronization has been attributed to play a role in movement \cite{Cassidy}, memory \cite{Klimesch} and epilepsy \cite{Lehnertz,Mormann}. It is clear that in all these examples the structure of the neural network must play a crucial role in its function. The adaptive development of the network structure takes place through the modifications of synaptic connections, governed by underlying neural learning mechanisms. Such synaptic modifications are posited to constitute the neural basis of learning and  the consequent acquisition of long term memory \cite{Abbott,Shimizu}. 

In the nervous system, a neuron integrates inputs from other neurons and generates outputs in the form of action potentials or spikes when its membrane potential exceeds an electrophysiological threshold. In particular, tonically spiking neurons are observed to ``fire'' spikes at regular intervals with a particular time period. Although the dynamics of single neurons are essential, complex cognitive phenomena emerge from the interactions of many neurons. In a given neuronal network, neurons that make synaptic connections influence one another through either excitation or inhibition. 

Collective synchronization in natural systems has been previously modeled by representing them as networks of coupled phase oscillators \cite{Cumin, Kori, Kuramoto, Maistrenko, Pazo, Tsimring}. These studies assumed a pre-imposed static network structure and connectivity. In particular, the influential Kuramoto Model \cite{Kuramoto} relied on global, all-to-all connectivity in which each oscillator affected every other oscillator equally. 

Recent theoretical efforts have studied how a network may develop in accordance with neural learning mechanisms in relation to the dynamics of synchronous cluster formation \cite{Aoki, Maistrenko, Ren, Seliger, Takahashi}. Neurophysiological studies have shown that a synapse is strengthened if the pre-synaptic neuron repeatedly causes the post-synaptic neuron to fire, leading to the Long Term Potentiation (LTP) of the synapse \cite{BlissLomo1973,Bliss}. Symmetrically, Long Term Depression (LTD) occurs when the post-synaptic neuron does not fire when the pre-synaptic neuron does. Experimental findings suggest further that learning may not depend solely on the rate of spikes at a synapse but on the relative timing of pre and post-synaptic spikes  \cite{BiPoo1, BiPoo2, Markram, Wittenberg}. According to the Hebbian theory \cite{Hebb}, the strength of the synapse between two neurons is enhanced if they are simultaneously coactive. In this work, we represent the relative time between spikes in the pre and post-synaptic neurons as the relative phase of a pair of coupled oscillators, and in this way the phase of an oscillator  may be used to represent the time between two spikes generated by a given tonically spiking neuron. The intrinsic frequency, the frequency of an oscillator independent of any influence from other oscillators, shall represent the natural firing-rate of a neuron in a network \cite{Hutcheon, Llinas}.

Phase oscillator models with slow time-varying coupling have previously been capable of displaying associative memory properties, while revealing parameter regimes for which both synchronized and unsynchronized clusters are stable \cite{Seliger, Maistrenko}. We explore how synchronization and learning mutually affect one another for both slow and fast learning rates. Similar recent models have assumed homogeneous networks with equal intrinsic frequencies \cite{Aoki}. We show, however, that an oscillator network develops stable synaptic couplings that depend on the relative intrinsic frequencies and on the learning rate, as well as on the initial network state. The paper is organized as follows: in Sec. \ref{sec:model} we introduce the model endowed with dynamic connectivity. In Sec. \ref{sec:FastLearning} and \ref{sec:SlowLearning} we focus  on the scenario when the network learns quickly and slowly, respectively. We extend our findings to the scenario when the network starts out without any connections and self-develops due to the mutual interaction of synchronization and learning in Sec. \ref{sec:selfdevelopment}. We summarize our findings and provide perspectives in Sec. \ref{sec:conclusion}.

\section{The Model}
\label{sec:model}
The Kuramoto Model \cite{Kuramoto} considers a system of limit-cycle oscillators moving through the phases of their cycles based on each oscillator's instrinsic frequency and its interaction with other oscillators: 

\begin{equation}\label{eq:Kuramoto}
\frac{d\phi_i}{dt}=\omega_{i}+\frac{1}{N}\displaystyle\sum_{j=1}^{N}K_{ij}F\left(\phi_j-\phi_i\right),
\end{equation}
where $\phi_i \in [0,2\pi)$ is the phase of the $i$th oscillator and $\omega_i$ is its intrinsic frequency.  The intrinsic frequencies $\omega_i$ can be drawn from a probability distribution $g(\omega)$, which is assumed to be unimodal and symmetrically distributed around its mean. $K$ is an $N$ x $N$ matrix of coupling coefficients, and $F$ is a coupling function with a period of $2\pi$. Following Kuramoto \cite{Kuramoto}, we assume $ F(\phi)=\sin(\phi)$. In order to measure the degree of synchronization, a global order parameter, $r$, is defined as

\begin{equation}\label{eq:orderofsync}
re^{i\psi}(t)=\frac{1}{N}\displaystyle\sum_{j=1}^{N}e^{i\phi_j(t)}.
\end{equation}

It represents the centroid of the phases with $r(t)$ corresponding to the coherence in phases and $\psi(t)$ respresenting the mean phase. Another convenient measure of synchronization is given by $r^2 \in [0,1]$, the square of the modulus of the order parameter.

If we assume constant and identical coupling coefficients, then $K_{ij}=K$ for all $i,j$. This is known as the globally coupled Kuramoto Model. Assuming such coupling, Eq. \eqref{eq:Kuramoto} becomes 

\begin{equation}\label{eq:GlobalKuramoto}
\frac{d\phi_i}{dt}=\omega_{i}+\frac{K}{N}\displaystyle\sum_{j=1}^{N}\sin(\phi_j-\phi_i).
\end{equation}

The Kuramoto Model then reduces to a mean-field model. Any particular oscillator is sensitive only to the mean, global properties of the entire system of oscillators, making the detailed configuration of coupled oscillators irrelevant. It can be shown \cite{KuraCrawford} that the degree of synchronization becomes nonzero (in a second-order phase transition) when $K>K_c$ where $K_c$ is a critical coupling. If the the distribution $g(\omega)$ of intrinsic frequencies is Gaussian, with a standard deviation $\sigma$, then

\begin{equation}\label{eq:KcGauss} 
K_c=\frac{2}{\pi g(0)} = \sqrt{\frac{8}{\pi}}\sigma.
\end{equation}

The system of coupled oscillators reaches an average degree of synchronization that is independent of initial conditions, whether the oscillators started out completely in phase or distributed over the unit circle  \cite{Lars}.

Instead of assuming a constant, pre-imposed connectivity and coupling matrix, we wish to investigate how this network develops through neural learning mechanisms and affects synchronous cluster formation, and vice versa. The learning mechanisms discussed above can be represented by dynamically varying coupling coefficients according to the rule

\begin{equation}
\frac{dK_{ij}}{dt}=\epsilon \left[G(\phi_i-\phi_j)-K_{ij}\right]. \label{Kuralearn}
\end{equation}

Choosing $G(\phi)=\alpha\cos(\phi)$ renders Eq. \eqref{Kuralearn} roughly equivalent to the Hebbian learning rule. Note that $\phi_i$ and $\phi_j$ are simultaneously coactive if they are in phase and hence representative of LTP. When they are in anti-phase, that is, $\phi_i-\phi_j=\pi$, the condition is representative of LTD. We define $\alpha$ to represents a \emph{Learning Enhancement factor}. It amplifies the amount of learning if two neurons are coactive. When the \emph{Learning Rate} $\epsilon$ is small, synaptic modification is slow. In this case, synchronized clusters are formed which are usually stable with respect to external noise \cite{Seliger}, although we discuss below how the stability depends on $\alpha$. Since such stabilization, as in the Hopfield model \cite{Hopfield}, is reflective of long-term associative memory formation \cite{HertzKroghPalmer}, such a representation would be expected to yield an important perspective on the mechanisms of learning. 

Combining the models of spontaneous synchronization and Hebbian learning, our joint dynamical system is represented by

\begin{equation}\label{eq:jointKuramoto}  
\frac{d\phi_i}{dt}=\omega_{i}+\frac{1}{N}\displaystyle\sum_{j=1}^{N}K_{ij}\sin\left(\phi_j-\phi_i\right) 
\end{equation}

\begin{equation}\label{eq:jointLearning}  
\frac{dK_{ij}}{dt}=\epsilon \left[\alpha\cos(\phi_i-\phi_j)-K_{ij}\right]. 
\end{equation}

The $K_{ij}$ in Eq. \eqref{eq:jointLearning} is a saturating term which prevents coupling coefficients from increasing or decreasing without bound. A hard bound such as restricting $|K_{ij}| \leq 1$ as in \cite{Ren} can effectively limit the steady state values of the coupling coefficients to one of the two hard bounds. Although such restrictions can account for the memorization of binary data, a soft bound such as the saturating term we enlist here can allow the network to possess more diverse connectivity. It should be noted that the number of parameters appearing in Eqs. \eqref{eq:jointKuramoto} and \eqref{eq:jointLearning} could be reduced by means of rescaling time and absorbing $\epsilon$. However, since both $\epsilon$ and $\alpha$ are meaningful from a neurophysiological perspective, we choose to leave the equations in their present form.

Whereas previous work focussed on slow learning \cite{Seliger, Maistrenko}, we explore the network's behavior for both fast and slow learning scenarios.  We observe qualitatively different behaviors depending on the values of $\epsilon$ and $\alpha$. Particularly, we observe that there is a critical value of the learning rate below which a single synchronous cluster is formed as in the original Kuramoto Model in Eq.  \eqref{eq:GlobalKuramoto}. Above this critical value, two synchronous clusters emerge (see Fig. \ref{fig:epsilonphasetrans}). We define a new order parameter, $r_2$, as

\begin{eqnarray}\label{eq:r2squared}
r^{'}e^{i\psi^{'}(t)} &=& \frac{1}{N}\displaystyle\sum_{j=1}^{N}e^{i2\phi_j(t)} \\ \nonumber
r_{2}^2 &=& |r^{'}-r|^2 
\end{eqnarray}
The subtraction of the degree of synchronization for a single cluster $r$ is necessary since $r^{'}$ is large also for the single cluster configuration; $r_2$ is designed to pick out the dipole moment of the distribution (i.e. two clusters). 

As we will show, the dynamics of the system depends on whether the learning rate parameter $\epsilon$ is large or small compared to some critical $\epsilon_c$. For fast learning, $\epsilon > \epsilon_c$, the coupling coefficients can adjust themselves rapidly enough according to Eq. (\ref{eq:jointLearning}) that they follow the ``fixed point'' $\alpha \cos(\phi_i-\phi_j)$ adiabatically as it oscillates before a synchronized state has manifested. For slow learning, $\epsilon < \epsilon_c$, the coupling coefficients cannot follow the oscillation, and thus they can only depart consistently from the initial values once a synchronized state has established itself.

To estimate the magnitude of $\epsilon_c$, we have to compare the rate at which Eq. (\ref{eq:jointLearning}) can change with the frequency of the term $\cos(\phi_i-\phi_j)$. It is clear that Eq. (\ref{eq:jointLearning}) would asymptotically approach a static fixed point with a time constant of $\tau_1=1/ \epsilon$. On the other hand, $\cos(\phi_i-\phi_j)$ is expected to oscillate at a frequency of $|\omega_i-\omega_j|$, and so on average $\tau_2=\pi / 2\sigma$ is the time it takes for two oscillators starting in phase to have moved $\pi/2$ out-of-phase where they do not influence their mutual coupling coefficient any longer. Setting these two time scales equal to one another yields, 

\begin{equation} \label{epscrit}
\epsilon_c= \frac{2\sigma}{\pi}. 
\end{equation}

In our study, $\sigma=0.1$, so that $\epsilon_c \approx 0.064$. It should be noted that the argument above only holds when the starting coupling coefficients of the network satisfy $K_{ij}(0)> K_c \approx 0.16$ (see Eq. (\ref{eq:KcGauss})). Below $K_c$, the system cannot attain global synchronization at all in the slow-learning regime.

\section{Fast Learning} \label{sec:FastLearning}
Situations where memorization of specific details are necessary involve fast learning. Hippocampal conjuctive coding in particular, is believed to involve such rapid, focused learning \cite{Mcclelland}. In our joint dynamical system, when $\epsilon>\epsilon_c$, the coupling coefficients can follow the ``fixed point'' and so $K_{ij}^*=\alpha \cos(\phi_i-\phi_j)$. Substituting $K_{ij}^*$ into Eq. \eqref{eq:jointKuramoto} then yields,
\begin{equation} \label{eq:largeepsilonKuramoto}
\frac{d\phi_i}{dt} = \omega_{i}+\frac{\alpha}{2N}\displaystyle\sum_{j=1}^{N}\sin\left[2(\phi_j-\phi_i)\right].
\end{equation}
Multiplying both sides of Eq. \eqref{eq:largeepsilonKuramoto} by 2 and defining ${\phi_i}^{'}=2\phi_i$ and ${\omega_i}^{'}=2\omega_i$ yields,

\begin{equation}\label{eq:largeepsilonKuramoto2}  
\dot{{\phi_i}^{'}}={\omega_i}^{'}+\frac{\alpha}{N}\displaystyle\sum_{j=1}^{N}\sin({\phi_j}^{'}-{\phi_i}^{'}). 
\end{equation}
This is equivalent to the global Kuramoto Model in Eq. \eqref{eq:GlobalKuramoto}, except that the phases are now in double angles. We would therefore expect to find a critical value of the learning enhancement factor, $\alpha_c$, at which a second-order phase transition to synchronization occurs. Under our previous assumptions for $g(\omega)$,

\begin{equation}\label{eq:pdfcomparison}  
\int_{-\infty} ^{\infty} g(\omega)d\omega=1=\int_{-\infty} ^{\infty} g^{'}(\omega ^{'})d\omega^{'},
\end{equation}
it follows that
\begin{equation}\label{eq:gvsg'}
g^{'}(\omega)=\frac{g(\omega)}{2}.
\end{equation} 
Accordingly, comparing with Eq. \eqref{eq:KcGauss},

\begin{equation}\label{alphac}
\alpha_c=\frac{2}{\frac{\pi g(0)}{2}}=2K_c
\end{equation}

\begin{figure}[h] 
   \centering
      \includegraphics[width=8cm]{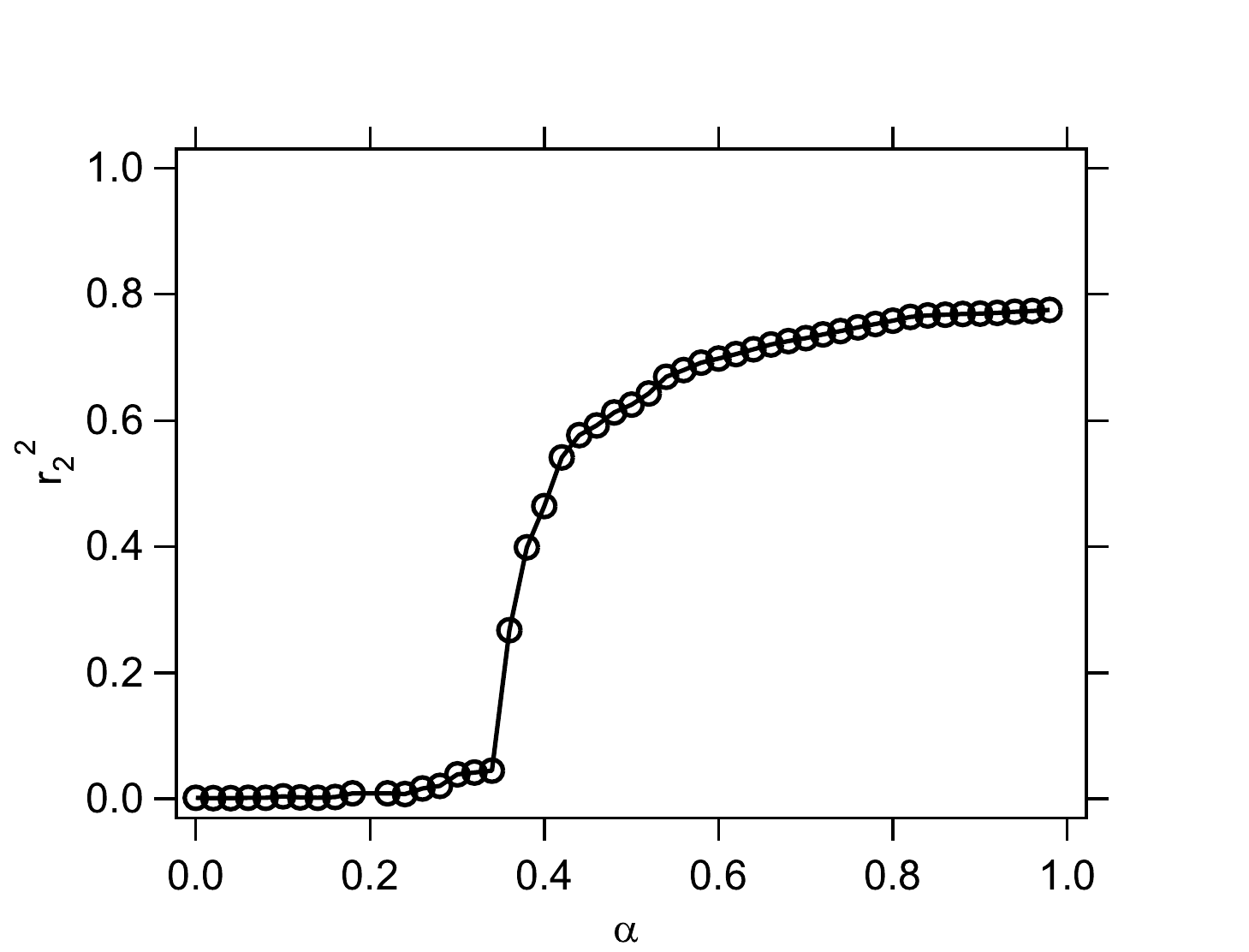} 
  \caption{The degree of synchronization ${r_2}^2$ as a function of the learning enhancement factor $\alpha$ when the initial phases of oscillators are uniformly distributed over the circle. $\epsilon>\epsilon_c$ was set at a large value of 1. A second-order phase transition is observed at $\alpha_c=2K_c=0.32$ for Gaussian intrinsic frequency distribution of standard deviation $\sigma=0.1$.}
   \label{fig:r2phasetrans}
\end{figure}

In order to verify the value of $\alpha_c$, we performed a series of numerical simulations. All simlulations in this study employed an Euler timestep of $\Delta t = 0.1$. In Fig. \ref{fig:r2phasetrans}, $\epsilon$ was set to 1.0, so that $\epsilon >\epsilon_c = 0.064$, and the network consisted of 500 oscillators. Intrinsic frequencies $\omega_i$ were drawn from a Gaussian distribution with mean $\mu=0$ and standard deviation $\sigma=0.1$. In this case, according to Eq. \eqref{eq:KcGauss}, $K_c \approx 0.16$. We then varied the value of $\alpha$ from 1 towards 0 and obtained a bifurcation diagram relating the average eventual degree of synchronization to the learning enhancement factor. We observe a second-order phase transition in $\alpha$ for the joint system similar to that of the original Kuramoto model with global all-to-all coupling. Critically, this phase transition occurs at $\alpha_c=0.32=2K_c$, verifying the theoretical prediction. This critical value is robust with respect to varying initial conditions of the phase distribution $\phi_i(0)$ and connectivity $K_{ij}(0)$.

\begin{figure}[htbp] 

  \includegraphics[width=9cm]{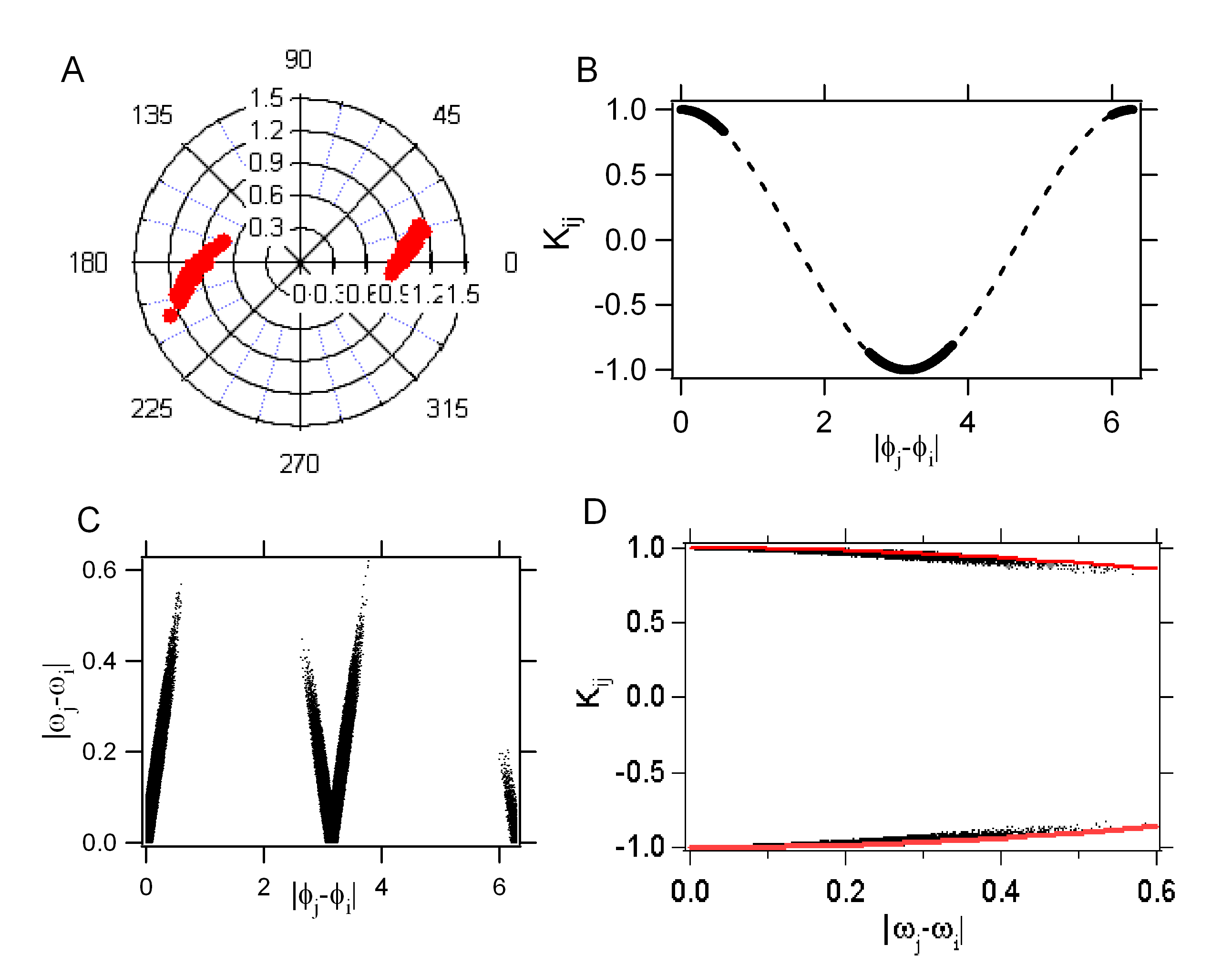} 
  \caption{Fast Learning ($\epsilon>\epsilon_c$) with $\alpha=1 >\alpha_c$. (A) Polar plot of the distribution of oscillators. Two stable clusters are formed. (B) Final $K_{ij}$ as a function of the final relative phases $|\Delta\phi_{ij}|$. The thick line represents the data from simulations, dashed curve is the theoretical prediction: $K_{ij}^*=\alpha \cos(\phi_i-\phi_j)$. (C) Relative intrinsic frequencies $|\Delta\omega_{ij}|$ as a function of final relative phases $|\Delta\phi_{ij}|$. (D) Final $K_{ij}$ as a function of relative intrinsic frequencies $|\Delta\omega_{ij}|$. Black dots represent data from simulations, red curve corresponds to the theoretical fit.}
   \label{fig:epsilon=1alpha=1}
 \end{figure}

When $\alpha>\alpha_c,\epsilon>\epsilon_c$, and all oscillators do not start out in phase, two clusters are formed, which remain 180 degrees apart from each other in mean phase as shown in Fig. \ref{fig:epsilon=1alpha=1}A. Analyzing the phase-plane we obtain four fixed points for the joint dynamics of Eqs. \eqref{eq:jointKuramoto} and \eqref{eq:jointLearning}. Two of them are stable, corresponding to $\phi_j-\phi_i=0$ and $\phi_j-\phi_i=\pi$. Thus, stable states for this system occur when pairs of oscillators are either synchronized or anti-synchronized with each other, leading to the formation of the two anti-synchronized clusters. The other steady states, corresponding to a relative phase of $\frac{\pi}{2}$ and $\frac{3\pi}{2}$, are unstable. It follows that for the stable steady states,
 
\begin{equation}\label{eq:Ksteady}
K_{ij}^*=\alpha \cos(\phi_i-\phi_j) \approx \pm\alpha
\end{equation}
with $K_{ij}^* \approx \alpha $ within a synchronized cluster and $K_{ij}^* \approx -\alpha $ between two anti-synchronized clusters. As seen in Fig. \ref{fig:epsilon=1alpha=1}B, the final steady-state values of the coupling coefficients for the two clusters, observed in the simulations, are in excellent accordance with the prediction of Eq. \eqref{eq:Ksteady}. The final values of the coupling coefficients $K_{ij}$ can also be correlated against the initial relative intrinsic frequencies of oscillators $\Delta \omega_{ij}=|\omega_{i}-\omega_{j}|$. Here it is useful to first relate the relative intrinsic frequencies of the oscillators to their final relative phases (Fig. \ref{fig:epsilon=1alpha=1}C). Within a cluster, we can calculate the slope of this relationship. The scatter-plot relating the final steady-state value of the coupling coefficients to the relative intrinsic frequencies of oscillators also depicts the formation of two clusters (Fig. \ref{fig:epsilon=1alpha=1}D). Using the slope of the line in Fig. \ref{fig:epsilon=1alpha=1}C together with the cosine fit in Fig. \ref{fig:epsilon=1alpha=1}B, we can again match the scatter plot very well.  

\begin{figure}[h] 
   \centering
   \includegraphics[width=7.5cm]{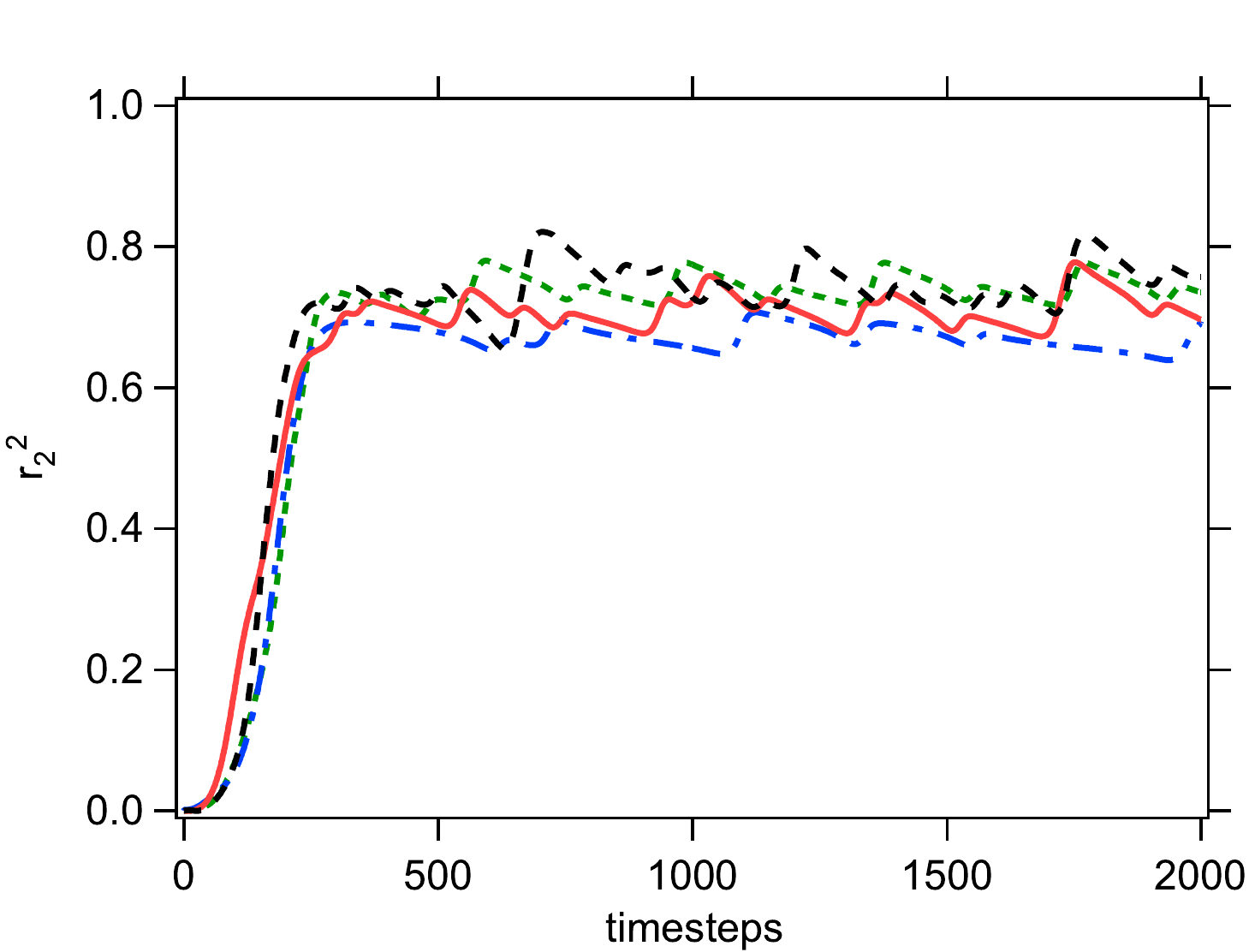} 
   \caption{The time evolution of the degree of synchronization $r_2$ when $\alpha>\alpha_c$ and $\epsilon$ is large (fast learning) for different initial coupling strengths $K(0)$. $\alpha$ was fixed at 0.5, $\epsilon=1$ and $K(0)$ was varied from 0, 0.25, 0.5 and 0.75. The degree of synchronization $r_2$ attained the same final value regardless of the initial coupling strengths.}
   \label{fig:r2icK}
\end{figure}

In the fast learning scenario, the strength of the initial network coupling has no effect on the eventual network structure (one or two clusters formed) or degree of synchronization. As shown in Fig. \ref{fig:r2icK}, regardless of whether we start the network without connections, with coupling coefficients $K(0)<\alpha$, $K(0)=\alpha$, or $K(0)>\alpha$, the degree of synchronization $r_2$ for two clusters always attains the same eventual value. In contrast to the original Kuramoto model of Eq. (\ref{eq:GlobalKuramoto}), here the degree of synchronization does depend on the initial relative phases of the oscillators and on the value of $\alpha$. When $\alpha>\alpha_c$ and all oscillators start out in phase, that is, $\phi_i(0)=0$ for $i=1,2,...N$, then only a single synchronized cluster is formed (the second cluster being viable but unpopulated), and a relatively large value of $r^2$ is observed, while $r_2^2$ tends toward 0. As discussed above, if $\alpha<\alpha_c$ then no synchronization can manifest.

\section{Slow Learning} \label{sec:SlowLearning}
In neuroscience, the ability of abstracting generalizable properties from specific details is believed to involve slow learning mediated by the neocortex \cite{Mcclelland}. In our model of coupled phase oscillators, slow learning occurs when $\epsilon<\epsilon_c$. Qualitatively, since there is very little change in Eq. \eqref{eq:jointLearning}, $K_{ij}\approx K_{ij}(0)=K>K_c$ on an intermediate time-scale. Substitution of this approximate condition into Eq. \eqref{eq:jointKuramoto} recovers the globally coupled Kuramoto Model given by Eq. \eqref{eq:GlobalKuramoto}. In this case, only a single synchronized cluster should form, and this result is easily verified by simultaions.

Whether this single cluster remains stable over long time scales depends on the value of $\alpha$. If $\alpha$ is chosen too low, an interesting phenomenon occurs whereby a cluster initially forms but at long times disintegrates again. The eventual disintegration is due to the decrease of the coupling coefficients at longer times below a value needed to sustain synchronization.

\begin{figure}[htbp] 
   \centering
   \includegraphics[width=8cm]{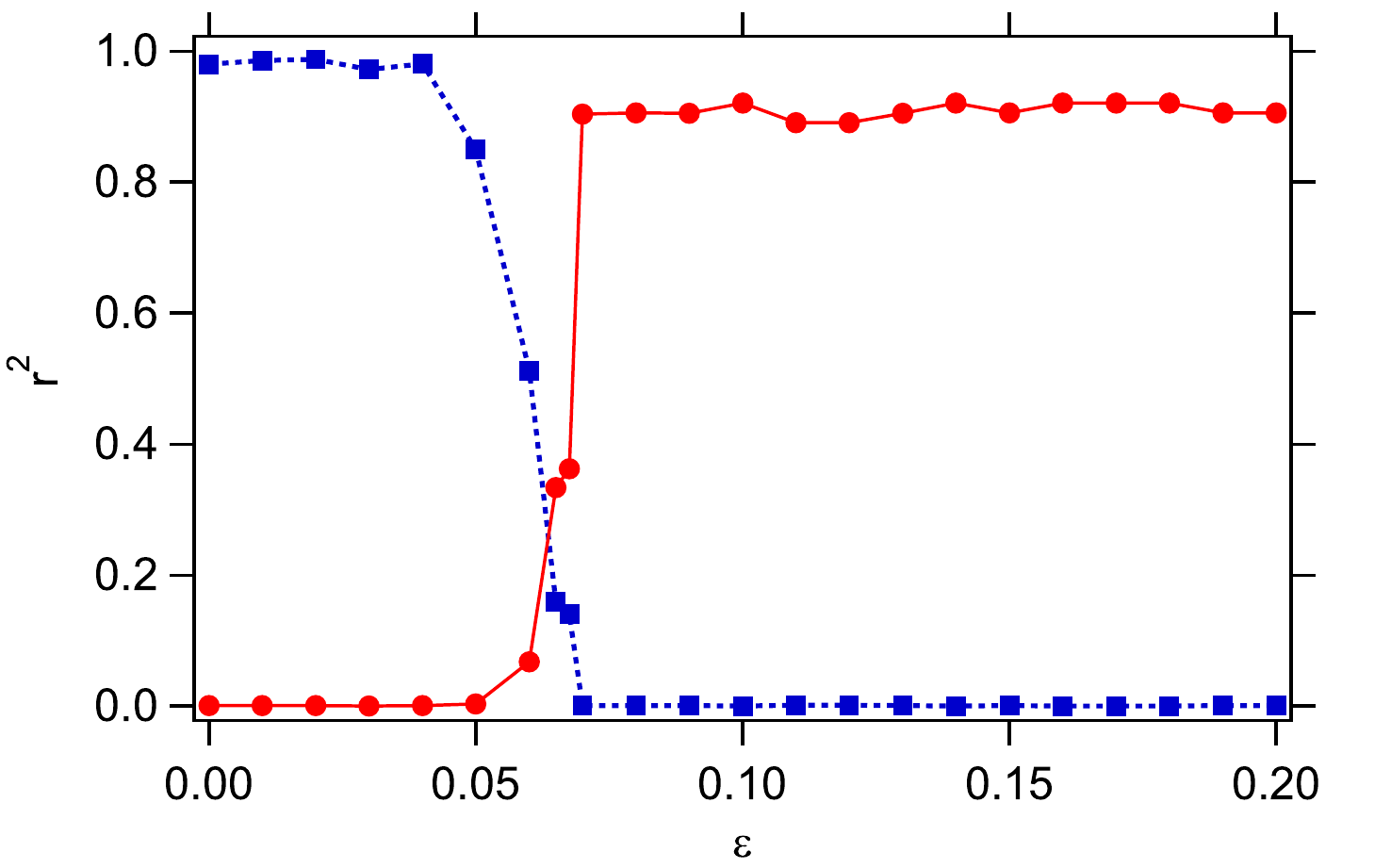} 
   \caption{The degree of synchronization a function of $\epsilon$ when $K(0)=0.75$. $\alpha>\alpha_c$ was set at a large value of 1. Blue and red traces represent the average eventual degree of synchronization for a single cluster $r^2$ and that for two clusters $r_2^2$, respectively. Each data point was computed by simulating a network of 250 oscillators for 5000 timesteps and averaging over the last 1000 timesteps.}
   \label{fig:epsilonphasetrans}
\end{figure}

Figure \ref{fig:epsilonphasetrans} summarizes the transition from a one-cluster state to a two-cluster state as $\epsilon$ is increased above the critical value. The blue trace depicts $r^2$ and the red trace ${r_2}^2$. The transition from a one-cluster state at small learning rates to a two-cluster state for fast learning is evident. A starting value of $K_{ij}(0)= 0.75 > K_c$ was used in the simuations shown, but other values of $K(0)$ were tested as well. Note that the transition between the one-cluster and two-cluster state occurs at the predicted $\epsilon_c=0.064$ separating slow and fast learning, thus veryfying the prediction of Eq. \eqref{epscrit}.

\section{Self-Development} \label{sec:selfdevelopment}
We now consider the intruiging case of $K_{ij}(0)=0$ for all $i,j$. This means that we start the joint system out with no connections between oscillators in order to observe how a connective structure may self-develop in this model. In neuroscience terms, we study whether parts of the nervous system can develop from the time of conception through the mutual interaction of spontaneous neural synchronization and Hebbian learning in order to perform their rich repertoire of functions. 

\begin{figure}[htbp] 
   \centering
   \includegraphics[width=8cm]{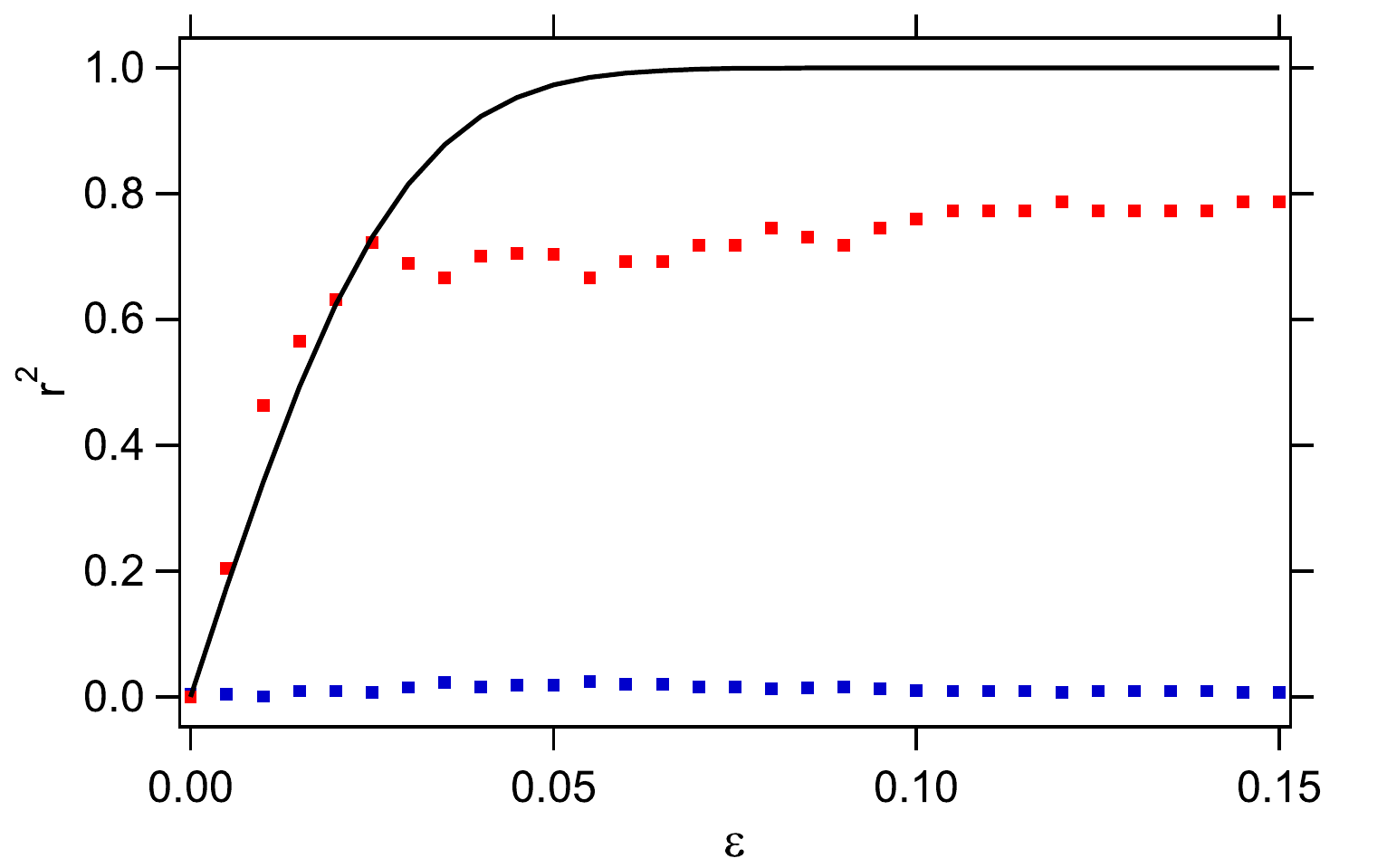} 
  \caption{Degree of synchronization ${r_2}^2$ as a function of learning rate $\epsilon$ when $K(0)=0$.  Red squares represent the numerical result for $r_2^2$ (computed as in Fig. \ref{fig:epsilonphasetrans}). The black line represents the predicted percentage of synchronized pairs. Blue dots represent the mean degree of sync for one cluster $r^2$.}
   \label{fig:selfassemblyepsilon}
\end{figure}

Let us investigate the role of the learning rate $\epsilon$ in the self-development of synchronized clusters. For this purpose, the learning enhancement factor, $\alpha$, is set to a value well above $\alpha_c = 2 K_c$ found earlier (see Fig. \ref{fig:r2phasetrans}). We would like to find the conditions that allow two oscillators that are near in phase at some instant of time to become entrained to one another. It is clear that in order for this to happen, 

\begin{eqnarray}
&\int_0^{T/4}{\dot{K} dt}\geq K_c, \mbox{with} \\ \nonumber
&K \approx \epsilon \alpha \cos(\Delta \omega t).
\end{eqnarray} 
$T$ denotes the time it would take for the unsynchronized oscillator-pair to diverge in phase by $2\pi$ and thus meet again; i.e. $T=2\pi / \Delta \omega$. Note that the distribution of frequency differences of oscillator-pairs is also normally distributed, but with a standard deviation increased by a factor of $\sqrt{2}$.

This condition implies that the two oscillators cannot be any further apart in intrinsic frequency than 
$\Delta \omega=\epsilon\alpha / K_c$. Thus, a first estimation of the percentage of oscillator pairs which are able to synchronize is given by the following function of $\epsilon$:

\begin{eqnarray}
&\int_{-\Delta\omega}^{\Delta\omega}{g(\omega)d\omega= \mbox{erf}(y)} \\ \nonumber
&y = \frac{\epsilon\alpha}{2\sigma K_c}
\end{eqnarray}

This relationship suggests that the degree of synchronization should depart roughly linearly from the origin as $\epsilon$ is raised from zero, indicating the absence of a phase transition in this case.  This prediction is confimred by numerical simulations. Figure \ref{fig:selfassemblyepsilon} shows the computed one- and two-cluster order parameters, $r^2$ and $r_2^2$, as a function of $\epsilon$. We observe that the one-cluster state does not occur for any value of $\epsilon$; it is `frozen' out for the initial condition $K(0)=0$. In contrast, the two-cluster state gradually turns on as $\epsilon$ is increased from zero.  

\begin{figure}[h] 
\centering
\includegraphics[width=7.5cm]{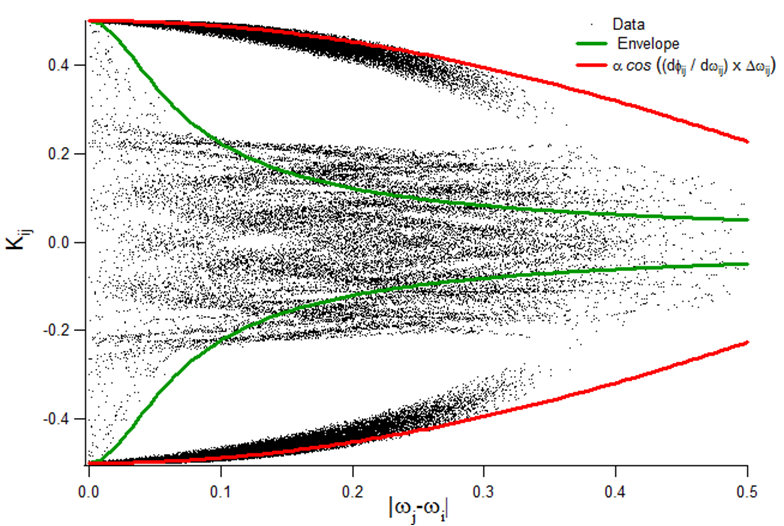} 
 \caption{Final coupling coefficients, $K_{ij}$, in a self-developed network plotted against $|\Delta\omega_{ij}|$. Black dots represent simulation results with $N=250$ oscillators, $\alpha=1,\epsilon=0.05$. Pairs of unsynchronized oscillators remain within an envelope while other oscillators form two synchronized clusters in anti-phase with each other.}
   \label{fig:Kzeroenvelope}
\end{figure}

In order to characterize the coupling coefficients that result from a self-assembled network further, let us examine the case $\epsilon=0.05$ (and $\alpha=1$, as before). Figure \ref{fig:Kzeroenvelope} depicts a scatter plot of the final coupling coefficients between all pairs of oscillators. We observe that the distribution falls into two distinct groups. The synchronized (and anti-synchronized) oscillator-pairs fall into the top and bottom arches. For the unsynchonized oscillators, an envelope (see green line in the figure) can be derived as follows:   

Since for this sub-population, $d\phi_i/dt \approx\omega_{i}$, after substition into Eq. \eqref{eq:jointLearning}, we obtain the first-order non-homogeneous differential equation 

\begin{equation}\label{eq:jointLearningKzero}  
\dot{K}_{ij} +\epsilon K_{ij}=\epsilon\alpha \cos(\phi_i-\phi_j)=\epsilon\alpha \cos(|\Delta \omega_{ij}|t),
\end{equation}
with solutions of 
\begin{equation}\label{eq:Lorentzian}  
K_{ij}(t)=\frac{\alpha}{\sqrt{1+({\frac{|\Delta \omega_{ij}|}{\epsilon}})^2}}\cos(|\Delta \omega_{ij}|t +\delta),
\end{equation}
where $\delta$ is a phase off-set. Thus, for unsynchronized oscillators, the relation between the coupling coefficients attained and the relative intrinsic frequencies remains within an envelope (Fig. \ref{fig:Kzeroenvelope}, green trace) that is expressed by the amplitude term in Eq. \eqref{eq:Lorentzian}.  For synchronized oscillators, the previous relationship holds (red-line).

\begin{figure}[h] 
 \centering
 \includegraphics[width=8.5cm]{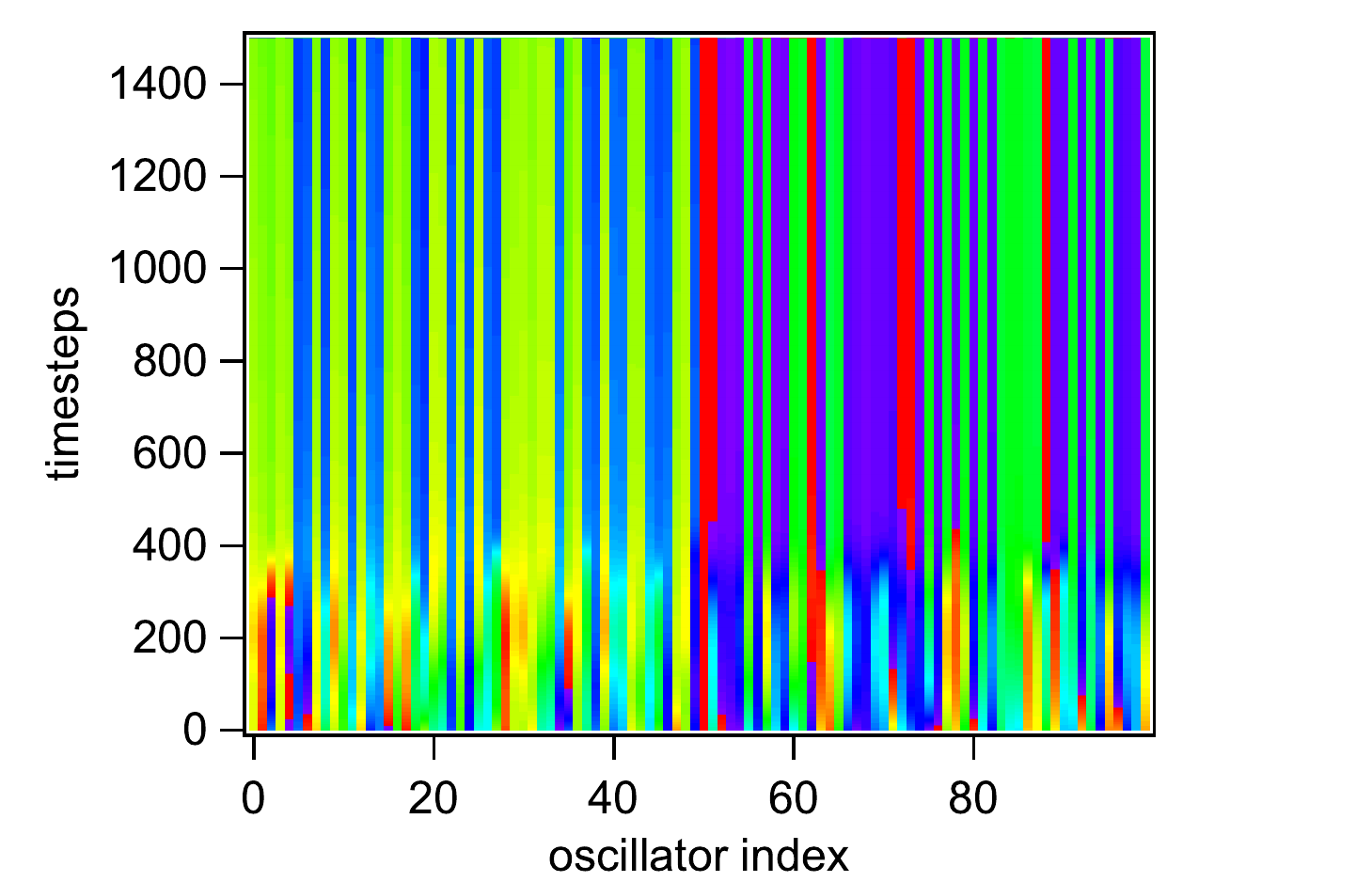} 
   \caption{The learning rate is abruptly changed from $\epsilon=0.1$ to $\epsilon=0.01$ at a timestep of 1000. The color indicates the oscillator phase relative to the middle oscillator (50). Once a stable pattern is established with fast learning, it does not change when the learning rate is reduced below $\epsilon_c$.}
   \label{fig:phaserel}
\end{figure}

Whereas  previous research demonstrated the ability of slow learning co-evolving networks to possess associative memory properties and learn binary patterns \cite{Seliger,Aoki}, here we provide a mechanism whereby the network generates and learns more diverse patterns even if learning is fast. Figure (\ref{fig:phaserel}) illustrates what happens when the learning rate $\epsilon$ is changed abruptly from a high to a low value. This discontinuity happens at $t=1000$ in the figure. The color in this density plot indicates the phase of the 100 oscillators relative to the middle one. We see that once fast learning establishes a stable pattern, the switch to slow learning does not alter this pattern. Stable learning of this kind has been put forward as a neurally plausible mechanism for the acquisition of long-term memories, especially with regard to the consolidation of declarative memories. According to the Complementary Learning Systems framework \cite{Mcclelland}, rapid, focused learning enables the acquisition of specific memories and their storage in the hippocampus, while the neocortex mediates the gradual extraction of structure and general properties, leading to the consolidation of these memories.

\section{Conclusion} \label{sec:conclusion}
In summary, we have explored the mutual effects of spontaneous synchronization and Hebbian learning in a neuronal network, focusing specifically on the role of the learning rate. Our work predicts qualitatively different behaviors of the network depending on whether learning is fast or slow. Specifically, unless the network is in a pre-existing state of phase synchrony, when learning is fast, it evolves into two anti-synchronized clusters as long as a learning enhancement factor, $\alpha$, is larger than a critical value. We found that $\alpha_c=2 K_c$. Furthermore, when learning is fast and $\alpha>\alpha_c$,the network always organizes itself into an all-to-all coupling structure with two clusters, regardless of its initial connectivity. Such anti-synchronized clustering is observed in neural systems known as Central Pattern Generators (CPGs) involved in mediating rhythmic motion \cite{KopellErmentrout,SigvardtWilliams,RandCohenHolmes,CruzCortez}. In these systems, when a group of neurons burst or fire in synchrony together, another group of neurons are inhibited to quiescence and vice versa. Whereas previous research has used phase oscillator models to account for such reciprocally inhibitory firing synchrony \cite{RandCohenHolmes,CruzCortez}, this study suggests how such networks may develop in the first place. 

We also predict a critical value in the learning rate, $\epsilon_c$, below which learning can be thought of as slow. In this regime, for sufficiently strong initial couplings, only one synchronized cluster forms. This synchronized cluster is stable and is maintained only if $\alpha$ is greater than its critical value. Otherwise, the network attains a state of synchrony but in the long-term returns to a state of disorder.

Finally, we extended our analysis to the case when a network starts out without any connections (or with sufficiently weak connections). We demonstrated that the degree of synchronization varies continuously with the learning rate and no phase transition is observed. Here when the learning rate is too slow, the network remains in an unsynchronized state indefinitely. Thus, our model predicts that if learning is too slow, a neuronal network cannot self-develop through the mutual interactions of neural synchronization and Hebbian learning. In such a case, pre-existing couplings, or pre-existing synapses, which are sufficiently strong, are necessary for the neuronal network to self-develop.


\begin{thebibliography}{99}

\bibitem{Buck} J. Buck, Quart. Rev. Biol. {\bf 63}, 265289 (1988).

\bibitem{BuckBuck}J. Buck and E. Buck, Sci. Am. {\bf 234}, 7485 (1976).

\bibitem{walker} T. J. Walker, Science {\bf 166}, 891 (1969).

\bibitem{michaels} D. Michaels, E. Matyas, and J. Jalife, Circulation Res. {\bf 61}, 704 (1987).

\bibitem{Sompolinsky1}
H. Sompolinsky, D. Golomb, and D. Kleinfeld, Phys. Rev. A {\bf 43}, 6990 (1991).

\bibitem{Galanetal}
R. F. Galan, N. Fourcaud-Trochme, G. B. Ermentrout, and N. Urban, J. Neurosc. {\bf26}, 3646 (1991).

\bibitem{KopellErmentrout}
N. Kopell, and G. B. Ermentrout, Comm. Pure and Appl. Math. {\bf 39}, 623 (1986).

\bibitem{SigvardtWilliams}
K. A. Sigvardt and T. L. Williams, Sem. in Neurosc. {\bf 4}, 37 (1992).

\bibitem{RandCohenHolmes}
R. H. Rand, A. H. Cohen, and P.J. Holmes, {\it Systems of coupled oscillators as models of central pattern generators}, in Neural Control of Rhythmic Behavior, 1988, (New York: Wiley), A.H. Cohen Eds.

\bibitem{CruzCortez}
F. Cruz and C. M. Cortez, Physica A {\bf 353}, 258 (2005). 

\bibitem{liu}
C. Liu, D. R. Weaver, S. H. Strogatz, and S. M. Reppert, Cell {\bf 91}, 855 (1997).

\bibitem{Cassidy}
M. Cassidy, et. al., Brain {\bf 125}, 1235 (2002). 

\bibitem{Klimesch}
W. Klimesch, Int. J. Psychophys. {\bf 24}, 61 (1996).

\bibitem{Lehnertz}
K. Lehnertz, Int. J. Psychophys. {\bf 34}, 45 (1999). 

\bibitem{Mormann}
F. Mormann, K. Lehnertz, P. David, and C. E. Elger, Physica D {\bf 144}, 358 (2000). 

\bibitem{Abbott}
L. F. Abbott, and S. B. Nelson, Nature Neurosc. {\bf 3}, 1178 (2000). 

\bibitem{Shimizu}
E. Shimizu, Y. P. Tang, C. Rampon, and J. Z. Tsien, Science {\bf 290}, 1170 (2000).

\bibitem{Cumin}
D. Cumin, and C.P. Unsworth, Physica D {\bf 226}, 181 (2007). 

\bibitem{Kori}
H. Kori, Y. Kuramoto, Phys. Rev. E {\bf 63}, 046214 (2001).

\bibitem{Kuramoto}
Y. Kuramoto, \emph{Chemical Oscillations, Waves and Turbulence}. 1984, (Berlin: Springer Verlag).

\bibitem{Maistrenko}
Y. L. Maistrenko, B. Lysyanski, C. Hauptmann, O. Burylko, and P. A. Tass, Phys. Rev. E {\bf 75}, 066207 (2007).

\bibitem{Pazo}
D. Pazo, Phys. Rev. E {\bf 72}, 046211 (2005).

\bibitem{Tsimring}
L. S. Tsimring, N. F. Rulkov, M. L. Larsen, and M. Gabbay, Phys. Rev. Lett. {\bf 95}, 014101 (2005).

\bibitem{Aoki}
T. Aoki and T. Aoyagi, Phys. Rev. Lett. {\bf 102}, 034101 (2009)

\bibitem{Ren}
Q. Ren and J. Zhao, Phys. Rev. E {\bf 76}, 016207 (2007).

\bibitem{Seliger}
P. Seliger, S. C. Young, and L. S. Tsimring, Phys. Rev. E {\bf 65}, 041906 (2002). 

\bibitem{Takahashi}
Y. K. Takahashi, H. Kori, and N. Masuda, Phys. Rev. E {\bf 79}, 051904 (2009).

\bibitem{BlissLomo1973}
T. V. P. Bliss and T. Lomo, J. Physiology {\bf 232}, 331 (1973).

\bibitem{Bliss}
T. V. P. Bliss and A. R. GardenerMedwin, J. Physiology {\bf 232}, 357 (1973).

\bibitem{BiPoo1}
G. Q. Bi and M. M. Poo, J. Neurosc. {\bf 18}, 10464 (1998).

\bibitem{BiPoo2}
G. Q. Bi and M. M. Poo, Ann. Rev. Neurosc. {\bf 24}, 139 (2001).

\bibitem{Markram}
H. Markram, et al., Science {\bf 275}, 213 (1997). 

\bibitem{Wittenberg}
G. M. Wittenberg and S. S. H. Wang, J. Neurosc. {\bf 26}, 6610 (2006).

\bibitem{Hebb}
D. O. Hebb, {\it The Organization of Behavior}. 1949, (New York: Wiley).

\bibitem{Hutcheon}
B. Hutcheon,and Y. Yarom, Trends in Neurosc. {\bf 23}, 216 (2000). 

\bibitem{Llinas}
R. R. Llinas, Science {\bf 242}, 1654 (1988).

\bibitem{KuraCrawford}
S. H. Strogatz, Physica D {\bf 143}, 1 (2000).

\bibitem{Lars}
L. Q. English, Eur. J. Phys. {\bf 29}, 143 (2008).

\bibitem{Hopfield}
J. J. Hopfield, Proc. Nat. Ac. Sc. U.S. - Bio. Sc. {\bf 79}, 2554 (1982).

\bibitem{HertzKroghPalmer}
J. Hertz, A. Krogh, and R. G. Palmer, \emph{Introduction to the Theory of Neural Computation}. 1991, (Boulder, CO: Westview Press).

\bibitem{Mcclelland}
J. L. McClelland, B. L. McNaughton, and R. C. O'Reilly, Psych. Rev. {\bf 102}, 419 (1995).


\end{thebibliography}
\end{document}